\documentclass[12pt]{article} 
	 
\usepackage{amsmath} 
\usepackage{psfrag,graphicx,verbatim,array,multicol,palatino,enumerate}
\usepackage{amsmath,amsfonts,bm,rotating,natbib}
\usepackage{pstricks,pst-node}
\usepackage{epsfig}
\usepackage{longtable}

\def\bmu{\boldsymbol{\mu}}
\def\bsigma{\boldsymbol{\sigma}}
\def\bTheta{\boldsymbol{\Theta}}

\def\bCV{{\rm {\bf{CV}}}}   
\def\bOV{{\rm {\bf{OV}}}}   
   
\def\bY{{\rm {\bf{Y}}}}   

\begin{document}

\begin{center} 
 
{\large{\bf EVALUATING THROWING ABILITY IN BASEBALL}}

\bigskip

MATTHEW CARRUTH\footnote{School of Engineering and Applied Science, University of Pennsylvania, Philadelphia, PA 19104  \\  {\tt{carruth@seas.upenn.edu}}}
and SHANE T. JENSEN\footnote{Department of Statistics, The Wharton School, University of Pennsylvania, Philadelphia, PA 19104  \\ {\tt{stjensen@wharton.upenn.edu}}}

\end{center}

\bigskip

\begin{abstract} 
We present a quantitative analysis of throwing ability for major league outfielders and catchers.   We use detailed game event data to tabulate success and failure events in outfielder and catcher throwing opportunities.  We attribute a run contribution to each success or failure which are tabulated for each player in each season.  We use four seasons of data to estimate the overall throwing ability of each player using a Bayesian hierarchical model.  This model allows us to shrink individual player estimates towards an overall population mean depending on the number of opportunities for each player.  We use the posterior distribution of player abilities from this model to identify players with significant positive and negative throwing contributions.   Our results for both outfielders\footnote{http://whartonball.blogspot.com/2007/03/evaluating-fielder-throwing-ability.html} and catchers\footnote{http://whartonball.blogspot.com/2007/04/evaluating-catcher-throwing-ability.html} are publicly available.
 
\end{abstract} 

\bigskip

\section{Introduction}\label{introduction}

The impact of a fielderÕs arm strength on their respective defensive rating has long been neglected and unmeasured.   Research into outfielder ability has tended to focus on estimation of differences in fielding range, such as the Ultimate Zone Rating \cite[]{Lic03} and the recent work by David Pinto \cite[]{Pin06}.  Despite this recent advancement in methods for quantifying the range of outfielders, there has been less development of sophisticated methods of quantifying an outfielderÕs throwing ability as a defensive tool.   The {\it put-outs} and {\it assists} statistics are a common but unacceptable summary of outfielder throwing ability since it only quantifies successful events, and outfielders are rarely given errors for unsuccessful ability to throw out players as a balancing measure of unsuccessful events.  In addition, there is a more subtle effect of throwing ability that is not captured by current measures: an outfielder with a reputation of a strong arm will be tested far less often and as a result, will save runs over the course of the season by reducing baserunner attempts to take extra bases.   More recent work has begun to address this need by quantifying both {\it hold} and {\it kill} events \cite[]{Wal07} but does not consider the influence of outfield ball-in-play location on these events.   

Research into catcher fielding also shows a lack of sophisticated analyses of throwing ability, which is even more necessary since there is limited information available for other aspects of catcher fielding.  Fielding range on pop-ups, bunts and short groundballs has been examined \cite[]{Pin06}, but these are relatively rare events, and in the case of pop-ups, the vast majority have large enough hang times that every catcher makes a successful play.  More success has been achieved in the study of passed balls and wild pitches, due to the work by David Gassko \cite[]{Gas05} and others.   Studies by Keith Woolner \cite[]{Woo99} have attempted to quantify differences between catchers in terms of pitch calling, but these effects have not been shown to be statistically significant.   Previous studies into throwing ability for catchers have not been satisfactory.   Most studies of throwing ability \cite[]{Tip97} have used broad categorizations such as caught stealing percentage, which does capture both successful and unsuccessful events to throw out baserunners.  However, the more subtle issue of attempt prevention is not captured by this statistic: a catcher with a reputation for high throwing ability will have less attempts to throw baserunners out, since baserunners will be less likely to attempt a stolen base.  The prevention of baserunning also has positive value to a team (though not as much as throwing out a baserunner), but this effect is not captured by statistics based only on baserunning attempts.  

In this paper, we quantify the relative throwing ability of each individual catchers and outfielders by tracking all throwing opportunities in the 2002-2005 seasons and evaluating the relative success of each fielder relative to the league average on the scale of runs saved/cost.  We capture both individual ability to throw out baserunners as well as individual tendencies to prevent baserunning.   We describe our overall methodology for catchers and outfielders in Section~\ref{methods}.  We focus on catchers and outfielders since the length of throws for infielders are so short that differences in throwing ability are impossible to separate from fielding ability.   In Section~\ref{shrinkage}, we average individual player throwing ability across multiple seasons with a simple hierarchical model (and Gibbs sampling implementation) that allows for different variances and sample sizes between different players.  We examine several interesting results from this approach for catchers in Section~\ref{catcherresults} and outfielders in Section~\ref{outfielderresults}.   We conclude with a brief discussion of our analysis.  

\section{Evaluation of Throwing Ability}\label{methods}

To track throwing events, we used the Baseball Info Solutions database \cite[]{bis07}  which tabulates detailed play-by-play data for all games within the 2002-2005 MLB seasons.   The BIS data contains all hitting and base-running events in each game, and any changes in scoring and baserunner configurations as a consequence of each event.  In addition, additional fielding information is provided for balls put into play: fielders involved in the play as well as the location where the ball was fielded. 
Our evaluation is based on an initial categorization of each event as a baserunning opportunity or not depending on the game situation.  In order to provide an interpretable and comparable measure for comparing individual players, we convert the successes or failures in these opportunities into a runs saved/cost measure.  The scale of runs saved/cost is a natural one and has been used previously by Gassko \cite[]{Gas05} in his work on the effects of passed balls and wild pitches.   Our overall strategy will be to apply the Expected Runs Matrix \cite[]{Rem07} to calculate the run contribution of a successful vs. unsuccessful plays, and reward/punish individual players accordingly.  These run rewards for individual players will be calculated while taking into account the averages across all players in order to come up with a runs saved/cost for each player.  

\subsection{Evaluation for Catchers}

Consider all events in the 2002-2005 seasons that involved a base-stealing opportunity.  For catchers, a successful play could either be catching a baserunner during a stolen base attempt, or preventing a baserunner from attempting a stolen base.  Steal opportunities  consisted of five categories: runner on first base, runner on second base, runners on first and third base, runners on second base with a runner on first,  and runner on first  base with a runner on second (a distinction is made between the last two categories in order to properly track double steals). These five  categories can be further divided into fifteen subcategories ($C = 1, ldots, 15$) based on how many outs there were prior to the play in question.   Steals of home plate are not considered by our analyses since they are usually the consequence of the pitcher, not of the catcher. 
For each catcher $P$, we tabulated all base-stealing opportunities $N(P,C)$ within each subcategory $C$ as well as the number of opportunities $A(P,C)$ where the baserunner did attempt a stolen base. If a stolen base was attempted, we tabulated the number of attempts that resulted in a successful steal $S(P,C)$ and the number of attempts where the baserunner was thrown out $F(P,C)$.  We also totaled these counts across all catchers in order to establish total values for the entire set of catchers.  An example of our tabulation is given in Table~\ref{tab-catcher} below. 
\begin{table}[ht]
\caption{Tabulation of Base-Stealing Opportunities for 2002}\label{tab-catcher}
\begin{center}
\begin{tabular}{|l|c|cccc|}
\hline
Catcher & Situation & Opportunities  & Attempts & Stolen & Caught  \\ 
 & $C$ &  $O_C$ & $A_C$ & Bases $S_C$ & Stealings $F_C$ \\ 
\hline
J. Lopez & man on 1st, 0 outs & 241 & 25 & 14 & 11  \\
All  &  man on 1st, 0 outs & 12361 & 831 & 519 & 312   \\
\hline
\end{tabular}
\end{center}
\end{table}
For each game situation $C$, we want to focus on two important concepts: the situation's propensity for steal attempts and the success rate of those steal attempts.  We use the totals over all catchers together with the opportunities for catcher $P$ to calculate expected counts of stolen bases and caught stealings for catcher $P$:  
\begin{eqnarray*}
{\rm E}[S(P,C)] = N(P,C) \times \frac{\sum\limits_{P} S(P,C)}{\sum\limits_{P} N(P,C)}  \hspace{1.5cm} 
{\rm E}[F(P,C)] = N(P,C) \times \frac{\sum\limits_{P} F(P,C)}{\sum\limits_{P} N(P,C)} 
\end{eqnarray*}
Returning to the situation in Table~\ref{tab-catcher} we expected J. Lopez to have ${\rm E}[S({\rm J.Lopez},C=1)] = 10.08$ stolen bases and ${\rm E}[F({\rm J.Lopez},C=1)]  = 6.07$ caught stealings.  Comparing to his observed totals, we see that  J. Lopez had 3.92 more stolen bases and 4.93 more caught stealings than expected in 2002.  How should J. Lopez be compensated for these individual differences in expectation for both stolen bases and caught stealings?   

The Expected Runs Matrix \cite[]{Rem07} provides us with expected runs for each game situation (baserunner configuration $\times$ number of outs), which we can use to calculate a runs saved/cost value for a particular successful or unsuccessful play.   As an example, consider again the game situation from Table~\ref{tab-catcher}.  From the Expected Runs Matrix, the expected run value $R$ for a (1st base alone, 0 outs) situation is 0.90.  If the baserunner attempts to steal and is thrown out by the catcher, then the game situation changes to (no baserunners, 1 outs) with a corresponding expected run value $R$ of 0.28, which means that the catcher has, in expectation, saved his team 0.62 runs.   However, if the baserunner successfully steals the base, then the game situation changes to (2nd base alone, 0 outs) with a corresponding expected run value $R$ of 1.14, meaning that the catcher has saved his team -0.24 expected runs (ie. cost his team 0.24 runs).   

More generally, for any game situation $C$ involving a baserunner, we have the run value $R(C)$ for that situation.  If a stolen base is attempted and a runner is caught stealing, the game situation changes from $C \longrightarrow C^\prime$, and the positive run value for the catcher is $R(C^\prime) - R(C)$.  If a stolen base is attempted and the runner gets a stolen base, the game situation changes from $C \longrightarrow C^{\prime\prime}$, and the negative run value for the catcher is $R(C^{\prime\prime}) - R(C)$.   Thus, the total number of runs saved by a catcher $P$ for a particular game situation $C$ is:
\begin{eqnarray}
\bCV (P, C) & = & \{F(P,C) - {\rm E}[F(P,C)]\} \times \{R(C^\prime) - R(C)\}  \nonumber \\
&& +   \{S(P,C) - {\rm E}[S(P,C)]\} \times \{R(C^{\prime\prime}) - R(C)\}  \label{equation1}
\end{eqnarray}
where $C^\prime$ is the change to situation $C$ from a caught stealing event, and $C^{\prime\prime}$ is the change to situation $C$ from a stolen base event.  Revisiting the example in Table~\ref{tab-catcher}, we see that in 2002, $\bCV({\rm J.Lopez}, C=1) = 4.93 \times 0.62 + 3.92 \times -0.24 = 2.12$ total runs saved in that game situation.   Equation (\ref{equation1}) must be evaluated for all fifteen game situations $C$, giving us a total runs saved/cost of 
\begin{eqnarray}
\bCV (P) & =&  \sum\limits_{C} \{F(P,C) - {\rm E}[F(P,C)]\} \times \{R(C^\prime) - R(C)\}  \nonumber \\
&& +   \{S(P,C) - {\rm E}[S(P,C)]\} \times \{R(C^{\prime\prime}) - R(C)\}
\label{equation2}
\end{eqnarray}
which is evaluated for each player in each year.    

\subsection{Evaluation for Outfielders}

The tabulation of outfielder throwing opportunities is somewhat more complicated than catchers.  We want to examine all ball-in-play (BIP) events to an outfielder that also had potential baserunning consequences.  For example, if a BIP event is a hit into the outfield, then it is a throwing opportunity only if there were baserunners on first and/or second base.   Hits with baserunners only on third base were not included since it is assumed that any baserunner can score from third base on a hit.  However, if the BIP event was an out (but not the third out), then the event can still be a throwing opportunity if there were baserunners on second and/or third base that could attempt to advance on the play.   We assume that baserunners will not advance from first base on an out unless there is another throwing event involved on the same play.   We categorize each outfielder throwing opportunity into a set of categories $C$ depending on the configuration of baserunners and whether the BIP was a hit or an out ($H=1$ for hit, $H=0$ for out).   In order to account for distance of the outfielder throw, the outfield surface was divided into a grid of 12 feet (X) by 10 feet (Y) zones $Z$, and each outfielder throwing opportunity was also categorized into a particular zone.  

Within each zone $Z$, and for every combination of baserunner configuration $C$ and hit vs. out $H$, we can tabulate the number of throwing opportunities $N(P,Z,C,H)$ for each player $P$.  We break these opportunities down into the number that resulted in thrown out baserunners $S(P,Z,C,H)$ and the number that resulted in runner advancements $F(P,Z,C,H)$.  Similar to our procedure for catchers,  we can compare the actual counts for outfielder $P$ to their expected counts based on their number of opportunities and the league totals:
\begin{eqnarray*}
{\rm E}[S(P,Z,C,H)] = N(P,Z,C,H) \cdot \frac{\sum\limits_{P} S(P,Z,C,H)}{\sum\limits_{P} N(P,Z,C,H)} \\
{\rm E}[F(P,Z,C,H)] = N(P,Z,C,H) \cdot \frac{\sum\limits_{P} F(P,Z,C,H)}{\sum\limits_{P} N(P,Z,C,H)}
\end{eqnarray*}
Similar to the catcher tabulation, we want to assign run values to the actions of an outfielder in each throwing opportunity, which again involves the use of the Expected Runs Matrix \cite[]{Rem07}.   For each throwing opportunity, we have the starting configuration of baserunners $C$, which has a certain run value $R(C)$.  If the throwing opportunity results in a thrown out baserunner, then our configuration changes to $C^\prime$ with run value $R(C^\prime)$, and the outfielder has contributed a positive run value of $R(C)-R(C^\prime)$.  However, if the throwing opportunity results in a runner advancement, then our configuration changes to $C^{\prime\prime}$ with run value $R(C^{\prime\prime})$,  then the outfielder has contributed a negative run value of $R(C)-R(C^{\prime\prime})$.  We can thus come up with the following total run contribution of an outfielder $P$ relative to the average:
\begin{eqnarray}
\bOV (P) & = &\sum\limits_{Z,C,H} \{S(P,Z,C,H) - {\rm E}[S(P,Z,C,H)]\} \times \{R(C)-R(C^\prime) \} \\ 
& + &\sum\limits_{Z,C,H} \{F(P,Z,C,H) - {\rm E}[F(P,Z,C,H)]\} \times \{R(C)-R(C^{\prime\prime})\}
\end{eqnarray}
This same evaluation was repeated for each player in each year.

\section{Hierarchical Model for Multiple Seasons}\label{shrinkage}

Our tabulation procedure in section~\ref{methods} gives us yearly throwing tabulations for 133 catchers and 500 outfielders.  In addition to evaluating catchers and outfielders on a seasonal basis, we also use these yearly totals to estimate the overall  throwing ability of each catcher and outfielder.  We use a simple hierarchical normal model designed to shares information between players while allowing for differences in variances between players.   We provide a short introduction to this model and also refer the reader to more detailed discussions in \cite{GelCarSte03}.   Consider the general situation of grouped data ie. $Y_{ij}$ where $j=1,\ldots,m_i$ indexes observations within group $i$ and $i = 1,\ldots,N$ indexes the groups.   We model our data as noisy observations centered at a group-specific mean $\mu_i$ and with a group-specific variance $\sigma^2_i$, 
\begin{eqnarray*}
Y_{ij} \sim {\rm Normal} (\mu_{i}, \sigma^2_i)  
\end{eqnarray*}
These group-specific means $\mu_i$ are also modeled as coming from a normal distribution,
\begin{eqnarray*}
\mu_i \sim {\rm Normal} (\mu_0, \tau^2) 
\end{eqnarray*}
The main parameters of interest are the unobserved means $\mu_i$ for each group $i$, which we will refer to collectively with the vector $\bmu$.  Inference for each $\mu_i$ is based on a balance between the observed mean $\overline{{\rm Y}}_i  = \sum_j Y_{ij} / m_i $ for that group and the population mean $\mu_0$ across all groups.   The details of that balance are determined by the within-player variance  and between-player variance  parameters, $\sigma^2_i$ and $\tau^2$.  Assuming that the parameters $\sigma^2_i$, $\tau^2$ and $\mu_0$ are known, then the best estimate of $\mu_i$ is:
\begin{eqnarray}
\hat{\mu}_i = \frac{\frac{m_i}{\sigma^2_i} \overline{{\rm Y}}_i  +  \frac{1}{\tau^2} \mu_0 }{\frac{m_i}{\sigma^2_i}   +  \frac{1}{\tau^2} }
\end{eqnarray}
A key consequence of this model is that the estimate $\hat{\mu}_i$ for a particular group is a compromise between the shared mean $\mu_0$ across groups and the group-specific mean of observed data $\overline{{\rm Y}}_i$.  The number of observations $m_i$ and amount of variance within the group $\sigma^2_i$ controls how much the resulting estimate $\hat{\mu}_i$ is {\it shrunk} towards the population mean $\mu_0$.  

In reality, the parameters $\sigma^2_i$, $\tau^2$ and $\mu_0$ are not known themselves. The Bayesian approach to this problem assumes prior distributions for these additional parameters.   Instead of focussing on a single point estimate of these parameters (such as the maximum likelihood estimate), we want to calculate the full posterior distribution of all unknown parameters  $\bTheta = (\bmu,\bsigma^2,\tau^2,\mu_0)$ given all observed data $\bY$.  Bayes rule is used to calculate this full posterior distribution: 
\begin{eqnarray}
p(\bTheta | \bY)  =  \frac{p(\bY | \bTheta) \cdot p(\bTheta)}{ p(\bY)}
\end{eqnarray}
The posterior distribution provides the entire range of reasonable values for our unknown parameters, but we need to summarize this distribution in a principled way.   We will use two summaries of each parameter in this paper: 
\begin{enumerate}
\item[a.] the posterior mean: $\hat{\mu}_i = {\rm E} (\mu_i | \bY_i )$ and 
\item[b.] the 95\% posterior interval: $(A,B)$ such that $P(A \leq \mu_i \leq B) = 0.95$.
\end{enumerate}
Posterior intervals are a similar concept to confidence intervals except that under the Bayesian approach, the parameter $\mu_i$ is considered to be a random variable taking the range of values given by the posterior interval.  In contrast, under the classical approach $\mu_i$ would be considered a fixed (but unknown) constant, and the range of values in a confidence interval refers to the coverage of this fixed constant across repeated samples.  Unfortunately, the posterior distribution $p(\bTheta | \bY)$ for this model is too complicated for posterior means and posterior intervals to be calculated analytically, and so we instead use a simulation-based approach called the Gibbs sampler to approximate the full posterior distribution  $p(\bTheta | \bY)$. The Gibbs sampler \cite{GemGem84}  samples values from the full posterior distribution $p(\bTheta | \bY)$ by iteratively sampling one parameter at a time from the conditional distribution of that parameter given the current values of all other parameters.  Specific details about the Gibbs sampler for our model are given in Appendix~\ref{modelimplementation}, and we again refer to  \cite{GelCarSte03} for a more involved discussion of Gibbs sampling and other simulation-based techniques.  

We now present the actual hierarchical model used for our analysis of catcher throwing ability, and the same model is also used for outfielder throwing ability.   In this application, the ``groups" described above represent individual players, and the observations within each group are the seasonal arm values for a particular player.  Our implemented model has the additional complication that we must also account for differences in the number of opportunities between player seasons.  Let $X_{ij}$ be the catcher run value $\bCV$ for catcher $i$ in season $j$, and let $n_{ij}$ be the number of opportunities for catcher $i$ in season $j$.   In order to compare different catchers on the same scale of opportunities, we calculate the average number of opportunities $\overline{n}$ in a season, and scale each run value by a factor of $n^\star_{ij} = n_{ij}/\overline{n}$: 
\begin{eqnarray*}
Y_{ij} = \frac{X_{ij}}{n^\star_{ij}} = X_{ij} \cdot \frac{\overline{n}}{n_{ij}}
\end{eqnarray*}
We model these re-scaled season run values as noisy observations from a underlying catcher-specific throwing talent $\mu_{i}$:
\begin{eqnarray}
Y_{ij} \sim {\rm Normal} (\mu_{i}, \sigma^2_i/n^\star_{ij})  \label{model1}
\end{eqnarray}
In addition to allowing player-specific variances $\sigma^2_i$ in model (\ref{model1}), we are also using $n^\star_{ij}$ to account for the fact that our observations $Y_{ij}$ should be more precise in seasons $j$ with greater numbers of opportunities $n_{ij}$.   The second level of our model allows information to be shared between catchers by assuming common distributions for the catcher-specific means $\bmu$ and variances $\bsigma^2$:
\begin{eqnarray}
\mu_i  \sim {\rm Normal} (\mu_0, \tau^2)  \qquad \qquad \sigma^2_i  \sim {\rm Inv-}\chi^2_\nu 
\end{eqnarray}
The parameters $\mu_0$ and $\tau^2$ capture the center and spread of the latent catcher-specific throwing abilities $\mu_i$.  As discussed before, we will use a Bayesian approach that treats our catcher-specific latent throwing abilities $\mu_i$ as random variables, and our inferential goal is the posterior distribution of each $\mu_i$.   As mentioned in our introduction to the normal hierarchical model, an important consequence of our model is that our resulting estimates of $\mu_i$ will be a compromise between the shared mean $\mu_0$ and the catcher-specific mean $\overline{Y}_i$ of scaled run values.  The amount of  {\it shrinkage} towards the shared mean for a particular $\mu_i$ will be a function of the catcher-specific variance $\sigma^2_i$ and the number of opportunities $n_{ij}$ for that player.   

To complete our model, we posit prior distributions $\mu_0 \sim {\rm N}(0,\beta) $ and $\tau^2 \sim {\rm Inv-}\chi^2_\gamma$.  Hyper-parameter ($\beta,\nu,\gamma$) values are used that make these prior assumptions {\it non-influential} on our inference, as discussed in Appendix~\ref{modelimplementation}.   In Section~\ref{catcherresults} below, we examine the results from our model implementation for catchers.  We also implemented this same model for our outfielder run values, and the results are given in Section~\ref{outfielderresults}.  
 
\section{Results for Catchers}\label{catcherresults}

\begin{figure}
\caption{95\% Posterior intervals for each catcher ordered by the posterior mean}\label{catcherpostint}
\begin{center}

\vspace{-1cm}

\rotatebox{270}{\includegraphics[height=6in,width=3.5in]{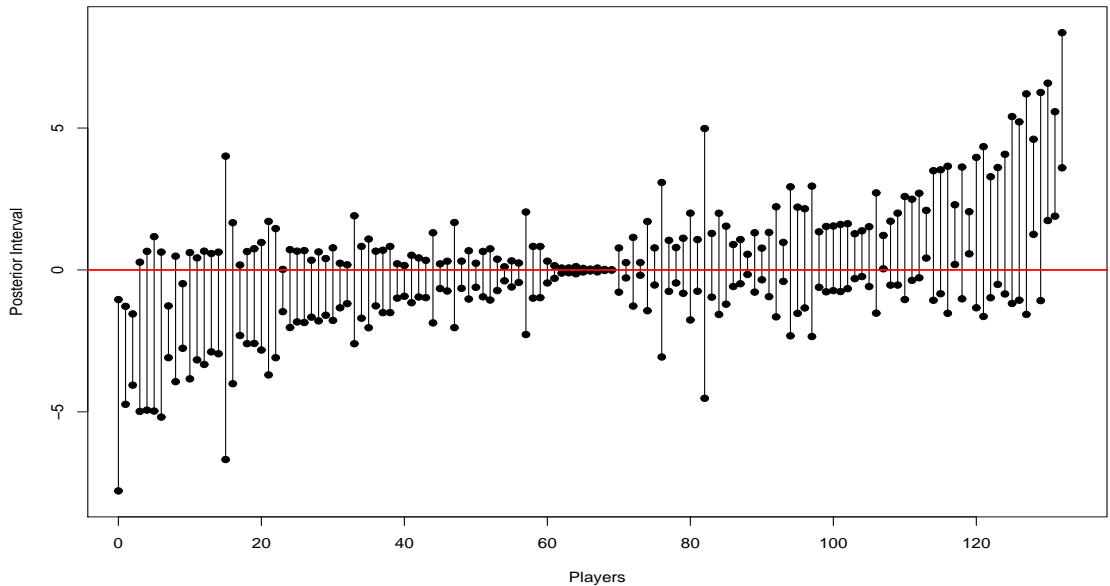}}
\end{center}
\end{figure}

We focus our inference on the marginal posterior distribution of each $\mu_i$, the latent throwing ability for each catcher $i$.    We calculated the posterior mean and 95\% posterior interval of $\mu_i$ for all 133 catchers in our dataset.   However, these $\mu_i$'s were estimated from our scaled run values $Y_{ij}$ that assumed the same number of opportunities for each catcher.  In order take into account differences in playing time, we also converted back to the scale of the original catcher-specific totals $X_{ij}$ by multiplying each posterior mean and posterior interval by the average number of opportunities for that catcher.   We use $\mu^\star_i$ to denote our re-scaled throwing contributions for each catcher $i$, which we call the player's {\it individual run contribution}.   For comparison, we call the scaled posterior mean $\mu_i$ the {\it scaled run contribution} for each player $i$.    Our posterior means and posterior intervals for the individual run contributions of  all 133 catchers are publicly available\footnote{http://whartonball.blogspot.com/2007/04/evaluating-catcher-throwing-ability.html}.  
\begin{table}[ht]
\caption{Catchers in 2002-05 with best and worst individual run contributions $\mu^\star_i$}\label{catchertable}
\begin{center}
\begin{tabular}{|lll|lll|}
\hline
\multicolumn{3}{|c|}{Five Best Catchers} & \multicolumn{3}{|c|}{Five Worst Catchers} \\
Name & Mean & Interval  & Name & Mean & Interval \\
\hline
Schneider, Brian &  6.55	 & (3.60, 8.36) &  Piazza, Mike & -4.98 & (-7.79, -1.05) \\
Molina, Yadier  & 4.41 & (1.89, 5.58) &  Varitek, Jason & -3.20 & (-4.74, -1.28) \\
Hall, Toby   & 4.23 & (1.74, 6.58) & Martinez, Victor & -3.00 & (-4.06, -1.55) \\
Ardoin, Danny  &  3.67 & (-1.08, 6.25) & Zaun, Gregg	& -2.78 & (-4.99, 0.27) \\
Miller, Damian  & 3.00 & (1.25, 4.61) & Fordyce, Brook & -2.68 & (-4.94, 0.66) \\
\hline
\end{tabular}
\end{center}
\end{table}
In Table~\ref{catchertable}, we give the five best and five worst catchers in terms of the posterior mean of their individual run contributions $\mu^\star_i$.  We also provide the 95\% posterior intervals for each of these catchers, and we observe that there is a large amount of variance.  This is not unexpected, considering that there are at most four seasons of observations for each catcher.  Even some of the best and worst catchers have posterior intervals that overlap with zero.   In Figure~\ref{catcherpostint}, we plot the 95\% posterior interval for all 109 catchers as a function of the posterior mean.  Only 13 of 133 catchers have 95\% posterior intervals that do not contain zero (indicated by the red line).  We also see that players with larger magnitudes of their run contributions also tend to have wider intervals for their individual run contributions.  

\section{Results for Outfielders}\label{outfielderresults}

Similar to our catcher analyses, we focus our outfielder inference on the {\it scaled run contribution} $\mu_i$ and {\it individual run contribution} $\mu^\star_i$ for each outfielder, which were calculated using the same methodology presented in Section~\ref{shrinkage}.   Just as before, the {\it scaled run contribution} $\mu_i$ are scaled to the same number of opportunities for each outfielder, whereas the {\it individual run contribution} $\mu^\star_i$ is re-scaled by the average number of opportunities faced by that particular outfielder $i$. Our posterior means and posterior intervals for the individual run contributions of  all outfielders are publicly available\footnote{http://whartonball.blogspot.com/2007/03/evaluating-fielder-throwing-ability.html}.  

\begin{table}[ht]
\caption{Outfielders in 2002-05 with best and worst individual run contributions $\mu^\star_i$}\label{outfieldertable}
\begin{center}
\begin{tabular}{|lll|lll|}
\hline
\multicolumn{3}{|c|}{Ten Best Outfielders} & \multicolumn{3}{|c|}{Ten Worst Outfielders} \\
Name & Mean & Interval  & Name & Mean & Interval \\
\hline
 Edmonds, Jim &  8.72 & (4.17, 12.40) &  Brown, Emil & -16.78 & (-20.82, -7.31) \\
 Jones, Jacque & 8.66 & (4.22, 12.78) &  Pierre, Juan & -10.77 & (-16.56, -4.06) \\
 Taveras, Willy & 8.07 & (0.21, 12.51) & Lawton, Matt & -9.43 & (-13.87, -4.29) \\
 Johnson, Kelly	& 7.77 & (1.98, 10.60) & Sanchez, Alex & -7.79 & (-12.27, -2.21) \\
 Sullivan, Cory	& 7.37 & (2.24, 10.36) & Holliday, Matthew	& -7.34& (-11.52, -1.81) \\
 Chavez, Endy	& 6.52 & (1.26, 10.80) & Crawford, Carl & -7.11 & (-10.60, -3.08) \\
 Guerrero, Vladimir & 6.20 & (-2.54, 13.66) & DeJesus, David & -7.00 &(-10.31, -3.17)\\
 Hidalgo, Richard & 6.18 & (-1.10, 11.92) & Williams, Bernie & -6.79 & (-11.28, -1.52)\\
 Hunter, Torii & 5.97 & (-0.21, 10.71) & White, Rondell & -6.56 & (-9.56, -2.76)\\
 Walker, Larry & 5.85 & (1.33, 9.32) & Magee, Wendell & -5.97 & (-9.47, -0.93) \\
 \hline
\end{tabular}
\end{center}
\end{table}

\begin{figure}
\caption{95\% Posterior intervals for each outfielder ordered by the posterior mean}\label{outfielderpostint}
\begin{center}

\vspace{-1cm}

\rotatebox{270}{\includegraphics[height=6in,width=3.5in]{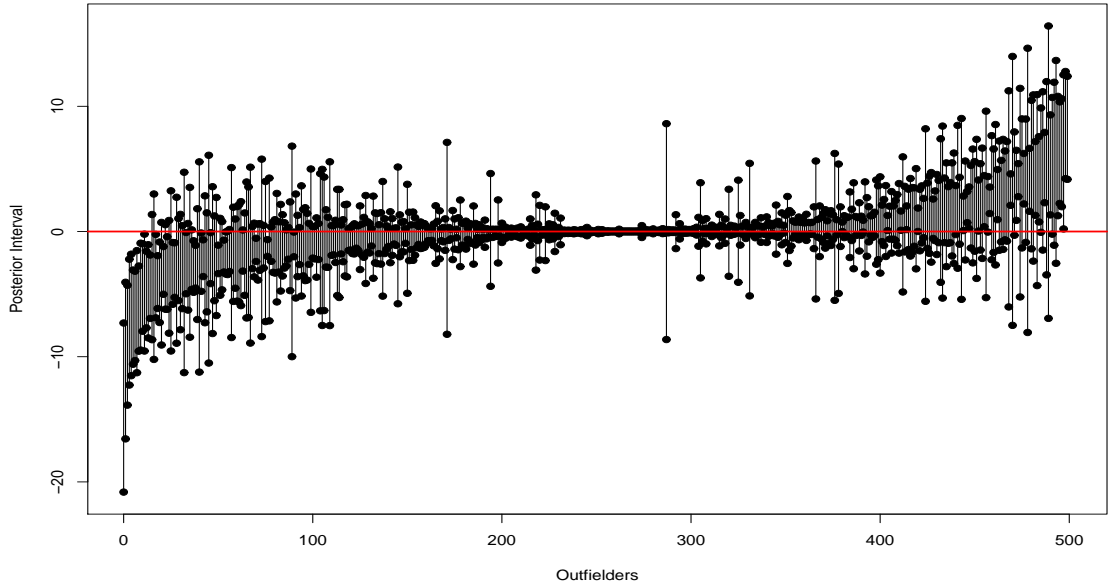}}
\end{center}
\end{figure}

In Table~\ref{outfieldertable}, we give the ten best and ten worst outfielders in terms of the posterior mean of their individual run contributions $\mu^\star_i$, along with 95\% posterior intervals.  We again observe a large amount of variance in the posterior intervals, and the magnitude of the run contribution for the best/worst outfielders is substantially greater than the magnitude of the run contribution for the best/worst catchers.    In Figure~\ref{outfielderpostint}, we plot the 95\% posterior interval for all 500 outfielders as a function of the posterior mean.  Only 60 of 500 catchers have 95\% posterior intervals that do not contain zero (indicated by the red line).  We again see that outfielders with larger magnitudes (highly positive or negative) of their run contributions also tend to have wider intervals for their individual run contributions.  

\section{Discussion}\label{discussion}

In this paper, we have presented an evaluation of throwing ability for both major league catchers and outfielders.  For outfielders, we focus on hits or outs on balls in play to the outfield when there are runners on base, whereas we focus our evaluation on base-stealing situations for catchers.   
For each player, our methodology tabulates the outcomes of their throwing successes and failures while taking into account the game situation for each opportunity.  As described in Section~\ref{methods}, we convert the performance of each player into runs saved/cost by tabulating the change in expected runs as a consequence of each of their throwing actions.    The run contribution for each player is calculated relative the average player, so a perfectly average defender would have a run contribution of zero.  The magnitude of run contributions is substantially higher for outfielders compared to catchers, which is a consequence of the greater number of throwing opportunities for outfielders as well as the generally greater run consequence of those opportunities.   

We use a hierarchical Bayesian model to estimate each player's innate throwing ability while accounting for differences between players in terms of variability and number of opportunities.   This model acts to shrink the run contribution of players with high variability or low numbers of opportunities towards the common mean of all players.  Based on this model, very few outfielders or catchers show significantly superior or inferior throwing performance, as defined by their 95\% posterior interval excluding zero.   The small number of statistically significant players is partly explained by the limitation of only having a maximum of four years of detailed game play data for each player.  Additional seasons of data for these players would reduce their posterior variance, which would likely lead to additional players with statistically significant abilities (i.e.  95\% posterior intervals excluding zero).  This additional data would also permit the extension of our hierarchical model to allow the throwing ability of individual players to change over time.  It would be difficult to model any time trends with our current four seasons of data, but this is a promising area of future research.

The use of the expected run matrix to evaluate the run consequence of stolen bases leads to an interesting consequence for our catcher evaluations.  In terms of change in expected runs, it is much more valuable for a catcher to throw out a baserunner that is attempting to steal than it is to prevent a baserunner from attempting a stolen base.  Some catchers that have a reputation for throwing out baserunners will not have as great of a run contribution because baserunners will attempt to steal less often, which is not rewarded as highly as throwing out a baserunner who does attempt a steal.  An example is Ivan Rodriguez who is considered to be the best catcher in the game, as evidenced by his 12 gold gloves, the most awarded to a individual catcher in the history of the award.  However, Rodriguez is ranked as only the eighth best catcher by our analyses, in part because he has one of the lowest proportion of steal attempts against him (3.47\% of baserunning opportunities) among regular catchers.   In Table~\ref{catchertable2}, we compare Ivan Rodriguez to Brian Schneider, the top MLB catcher by our analysis.   We see that Brian Schneider's overall run contribution is aided by the fact that he has a higher attempt percentage (4.54\% of baserunning opportunities) relative to Rodriguez.  Clearly, the optimal situation for a catcher is to have a high success rate on throwing out baserunners but without the reputation for doing so, so that baserunners still attempt to steal at a substantial rate.   An extreme (and not recommended) implementation of this strategy would suggest that catchers could deliberately fail on a throwing attempt in an relatively unimportant game situation in the hopes that baserunners would then be more likely to attempt (and be thrown out) in a more important situation.    \begin{table}[ht]
\caption{Comparison of Brian Schneider and Ivan Rodriguez}\label{catchertable2}
\begin{center}
\begin{tabular}{|llll|}
\hline
Name & Attempt \% & Mean & Interval  \\
\hline
Schneider, Brian & 4.54 \% &  6.55 & (3.60, 8.36)  \\
Rodriguez, Ivan  & 3.47\%  & 2.37 & (-1.18, 5.41) \\
\hline
\end{tabular}
\end{center}
\end{table}

Considering further the issue of situational importance, a potential extension of our model would be the incorporation of some measure of leverage into the tabulation of throwing events.  One could argue that catchers or outfielders should be rewarded more for making a successful throw or penalized more for failure in a key situation.  It is not clear, however, whether the innate throwing ability that we are attempting to capture with our method should be dependent on the importance of the situation.  The question of whether high leverage situations lead to a measurable difference in the performance of individual baseball players is a subject of ongoing speculation (eg. \cite{Tan04}).  

\section*{Acknowledgments}

The authors thank Abraham Wyner and Dylan Small for helpful discussion and suggestions. 

\bibliography{references}

\begin{appendix}

\section{Hierarchical Model Implementation}\label{modelimplementation}

As outlined in Section~\ref{shrinkage}, we have an observed number of opportunities $n_{ij}$ and run value $X_{ij}$ for catcher $i$ in season $j$.  Although the following model implementation is presented for our catcher evaluations, we use the same methodology for our outfielder analyses.  We scale our run values $X_{ij}$ to be on the same scale of opportunities, 
\begin{eqnarray*}
Y_{ij} = \frac{X_{ij}}{n^\star_{ij}} = X_{ij} \cdot \frac{\overline{n}}{n_{ij}}
\end{eqnarray*}
and then model these scaled run values as 
\begin{eqnarray}
Y_{ij} \sim {\rm Normal} (\mu_{i}, \sigma^2_i/n^\star_{ij})  \label{model1appendix}
\end{eqnarray}
where the parameters of interest are the underlying catcher-specific throwing talent $\mu_{i}$.  We share information between catchers by assuming common distributions for the catcher-specific means $\bmu$ and variances $\bsigma^2$:
\begin{eqnarray}
\mu_i  \sim {\rm Normal} (\mu_0, \tau^2)  \qquad \qquad \sigma^2_i  \sim {\rm Inv-}\chi^2_\nu 
\end{eqnarray}
Finally, we have the following prior distributions for our common mean $\mu_0$ and variance $\tau^2$:
\begin{eqnarray}
\mu_0  \sim {\rm Normal}(0,\beta) \qquad \qquad \tau^2 \sim {\rm Inverse-}\chi^2_\gamma 
\end{eqnarray}
The hyper-parameters $\nu,\beta,$ and $\gamma$ are assumed to be fixed and known, which means that we need to calculate the posterior distribution of our remaining unknown parameters $\bTheta = (\bmu,\bsigma^2,\mu_0,\tau^2)$, 
\begin{eqnarray}
p(\bTheta | \bY) & \propto & \prod\limits_{i=1}^N \prod\limits_{j=1}^{m_i} p(Y_{ij} | \mu_i, \sigma_i^2, n_{ij}) \cdot p(\mu_i | \mu_0, \tau^2) \cdot p(\sigma_i^2 | \nu) \cdot p(\mu_0 | \beta) \cdot p(\tau^2 | \gamma) \nonumber \\
 & \propto &  \prod\limits_{i=1}^N \left[ (\sigma_i^2)^{- \left(\frac{m_i+\nu}{2} + 1 \right)} \exp \left(- \sum\limits_{i=1}^N  \frac{1}{2 \sigma^2_i}\sum\limits_{j=1}^{m_i} (n^\star_{ij} (Y_{ij} - \mu_i)^2 + 1) \right) \right] \nonumber \\ 
& \times & (\tau^2)^{- \left(\frac{m_i}{2} + \frac{\gamma}{2} +1 \right)} 
\exp \left(\frac{-1}{2 \tau^2} (\sum\limits_{i=1}^N (\mu_{i} - \mu_0)^2 + 1) + \frac{1}{2\beta} \mu_0^2 \right) 
\end{eqnarray}
where $N$ is the number of catchers and $m_i$ is the number of seasons with a non-zero number of opportunities for catcher $i$.  We will estimate this posterior distribution with a Gibbs sampling strategy \cite[]{GemGem84} which consists of iteratively sampling from the following conditional distributions:
\begin{enumerate}
\item $p(\bmu | \bsigma^2, \mu_0, \tau^2, \bY)$
\item $p(\bsigma^2 | \bmu, \mu_0, \tau^2, \bY)$
\item $p(\mu_0 | \bsigma^2, \bmu, \tau^2, \bY)$
\item $p(\tau^2 | \mu_0, \bsigma^2, \bmu, \bY)$
\end{enumerate}
Step 1 can be done individually for each $\mu_i$.  The conditional distribution of each $\mu_i$ given the other parameters is 
\begin{eqnarray}
\mu_i \sim {\rm Normal} \left( \frac{\frac{1}{\sigma_i^2} \sum\limits_{j}^{m_i} n^\star_{ij} Y_{ij} +  \frac{1}{\tau^2} \mu_0}{\frac{1}{\sigma_i^2} \sum\limits_{j}^{m_i} n^\star_{ij} +  \frac{1}{\tau^2}} \, , \, \frac{1}{\frac{1}{\sigma_i^2} \sum\limits_{j}^{m_i} n^\star_{ij} +  \frac{1}{\tau^2}} \right)
\end{eqnarray}
We see that each catcher-specific throwing talent $\mu_i$ is a weighted compromise between the observed data $\sum_{j}^{m_i} n^\star_{ij} Y_{ij}$ and the common mean $\mu_0$.  The amount of {\it shrinkage} towards this common mean $\mu_0$ for a particular catcher is based on their number of opportunities and variance $\sigma_{i}^2$.  Catchers with low variance $\sigma_{i}^2$ will not be pulled as much towards the common mean $\bmu_0$.  

Step 2 can be done individually for each $\sigma_i^2$.  The conditional distribution of each $\sigma_i^2$ given the other parameters is 
\begin{eqnarray}
\sigma_i^2 \sim {\rm InverseGamma} \left( \frac{m_i + \nu}{2} \, , \, \frac{\sum\limits_{j=1}^{m_i} (n^\star_{ij} (Y_{ij} - \mu_i)^2 + 1}{2} \right)  \label{sigcond}
\end{eqnarray}
From equation (\ref{sigcond}), we see that our prior distribution on $\sigma_i^2$ are made non-influential relative to the data by letting $\nu \rightarrow 0$.  

For step 3, we use the following conditional distribution of $\mu_0$ given the other parameters, 
\begin{eqnarray}
\mu_0 \sim {\rm Normal} \left( \frac{\frac{N}{\tau^2} \overline{\mu}}{\frac{N}{\tau^2} + \frac{1}{\beta}} \, , \,  \frac{1}{\frac{N}{\tau^2} + \frac{1}{\beta}} \right) \label{mucond}
\end{eqnarray}
From equation (\ref{mucond}), we see that our prior distribution on $\mu_0$ are made non-influential relative to the data by letting $\beta \rightarrow \infty$.  

For step 4, we use the following conditional distribution of $\tau^2$ given the other parameters, 
\begin{eqnarray}
\tau^2 \sim {\rm InverseGamma} \left( \frac{N + \gamma}{2} \, , \, \frac{\sum\limits_{i=1}^{N} (\mu_{i} - \mu_0)^2 + 1}{2} \right)  \label{taucond}
\end{eqnarray}
From equation (\ref{taucond}), we see that our prior distribution on $\tau^2$ are made non-influential relative to the data by letting $\gamma\rightarrow 0$.  

\end{appendix}

\end{document}